\documentclass[aps,prl,twocolumn,showpacs,preprintnumbers]{revtex4}
\usepackage{graphicx}
\usepackage{amssymb}
\usepackage{amsmath}
\usepackage{times}
\usepackage{latexsym}
\def\beq{\begin{equation}}
\def\eeq{\end{equation}}
\def\bey{\begin{eqnarray}}
\def\eey{\end{eqnarray}}

\def\lsim{\mathrel{\raise.3ex\hbox{$<$\kern-.75em\lower1ex\hbox{$\sim$}}}}
\def\gsim{\mathrel{\raise.3ex\hbox{$>$\kern-.75em\lower1ex\hbox{$\sim$}}}}

\newcommand{\be}{\begin{equation}}
\newcommand{\ee}{\end{equation}}

\newcommand{\mr}[1]{\mathrm{#1} }
\begin{document}

\preprint{MCTP/09-45}

\title{Leptophilic Dark Matter from the Lepton Asymmetry}
\author{Timothy Cohen$^1$ and Kathryn M. Zurek$^{1,2}$}
\address{$^1$Michigan Center for Theoretical Physics, University of Michigan, Ann Arbor, MI 48109  \\ $^2$Particle Astrophysics Center, Fermi National Accelerator Laboratory, Batavia, IL  60510
}

\date{\today}

\begin{abstract}

We present a model of weak scale Dark Matter (DM) where the thermal DM density is set by the lepton asymmetry due to the presence of higher dimension lepton violating operators.  In these models there is generically a separation between the annihilation cross-section responsible for the relic abundance (through lepton violating operators) and the annihilation cross-section that is relevant for the indirect detection of DM (through lepton preserving operators).   Due to this separation, there is a perceived boost in the annihilation cross-section in the galaxy today relative to that derived for canonical thermal freeze-out.  This results in a natural explanation for the observed cosmic ray electron and positron excesses, without resorting to a Sommerfeld enhancement.  Generating the indirect signals also sets the magnitude of the direct detection cross-section which implies a signal for the next generation of experiments.  More generically these models motivate continued searches for DM with apparently non-thermal annihilation cross-sections.  The DM may also play a role in radiatively generating Majorana neutrino masses.

\end{abstract}
\maketitle

In recent decades a canonical model for Dark Matter (DM) utilizing the existence of Weakly Interacting Massive Particles (WIMPs) has emerged.   In models which stabilize the Higgs mass at the electroweak scale, the lightest of the new states introduced in these theories is often ``accidentally" stable due to a symmetry which is imposed for other reasons, such as $R$-parity.  The observed DM density, set by thermal freeze-out, determines the cross-section to annihilate to Standard Model (SM) fields to be a value typical of weak scale physics, $\langle \sigma\, v\rangle \simeq 3 \times 10^{-26} \mbox{ cm}^3/\mbox{s}$.  Within the paradigm of these models, many phenomenological expectations have been fixed, including the annihilation modes to the SM interaction channels with corresponding rates for indirect detection in the galaxy today.

However, the phenomenological successes of thermal WIMP DM can be preserved in other paradigms.  For example, the lepton or baryon asymmetry may set the DM density \cite{abunchofrefs}.  In these so called Asymmetric Dark Matter (ADM) models \cite{ADM}, DM in the GeV-TeV mass scale range naturally generates the observed relic abundance without standard thermal freeze-out.  When the DM from these models is hidden ({\em i.e.} it carries no SM charges) \cite{ADM}, its interactions with the SM fields may be set by interactions with new messengers (as in a Hidden Valley \cite{HV}) rather than with the SM electroweak fields or their superpartners.  And since the DM density is set by the lepton or baryon asymmetry, the SM-DM interactions are typically leptophilic or baryophilic respectively.  In addition, because the relic density is not set by the usual thermal freeze-out calculation, the relation between the DM density and the annihilation cross-sections relevant for the indirect observation of the DM today is modified.

Recent observations provide additional motivation for studying these models.  An excess in cosmic ray positron and electron signals over the expected background as observed by AMS-01 \cite{AMS}, HEAT \cite{Heat}, PPB-BETS \cite{PPB-BETS}, PAMELA \cite{Pamela}, Fermi \cite{fermi} and ATIC \cite{ATIC} may be a signal of annihilating DM.  The annihilation cross-section needed to produce these signals is non-thermal, a factor $\sim10-1000$ (depending on DM mass and astrophysical boost factor) larger than the thermal annihilation cross-section \cite{cholis,Strumia}.  Annihilation predominantly to leptons is preferred both by the shape of the PAMELA signal and the lack of excess in the anti-proton data \cite{salati}.  These facts appear to disfavor an explanation utilizing a canonical neutralino (though when combined with an astrophysical flux, it may be obtained \cite{michigan}).  One possibility is to introduce new GeV scale particles \cite{ArkaniWeiner}.  These light states mediate a Sommerfeld enhancement \cite{sommerfeld}, implying boosted annihilation in the halo today, while also acting as intermediate final states, thereby providing kinematic constraints on the allowed SM particles produced from DM annihilations.

In this letter we provide a simple paradigm which gives rise to both boosted and leptophilic annihilation of DM, involving neither Sommerfeld enhancements nor new GeV mass states.  When the DM relic density is set by the lepton asymmetry, the annihilation modes are naturally leptophilic.  Additionally, this density is derived using lepton number ($L$) violating operators that transfer the asymmetry, and not the $L$-preserving operators which lead to a signal for indirect detection experiments (such as PAMELA and Fermi) at low temperatures \footnote{Previous works considered DM from the lepton asymmetry as an explanation of the cosmic ray positron excesses, but utilized decaying DM with a lifetime tuned to the age of the universe \cite{decay}.  There have also been other models of leptophilic DM unrelated to the lepton asymmetry that have utilized a Sommerfeld enhancement to generate the boost \cite{leptophilic}}.  Though these models can provide a unique explanation for the cosmic ray excesses, their interest extends beyond this application.

We begin by outlining the general features of this class of models and then turn to constructing a simple model for illustration.  An initial lepton asymmetry is generated at temperatures well above the electroweak scale.  We are agnostic about the source of this asymmetry for the purposes of this paper.  Lepton number violating operators, which connect the SM leptons to dark sector fields, transfer the lepton asymmetry to the dark sector.  As in all models of ADM, these operators relate the DM number density to the lepton, and therefore baryon, density,
\begin{equation}
(n_X-n_{\bar{X}}) \sim (n_{\ell} - n_{\bar{\ell}})\sim (n_b-n_{\bar{b}}),
\end{equation}
where the exact proportions are ${\cal O}(1)$ and are determined by the particular operator transferring the asymmetries, and $(n_X - n_{\bar{X}})$, $(n_{\ell} - n_{\bar{\ell}})$ and $(n_b-n_{\bar{b}})$ are the asymmetries in the DM ($X$), leptons and baryons respectively.  As a result $m_X \sim \frac{\Omega_\mr{DM}}{\Omega_b}\, m_p$, where $m_X$ is the DM mass, $m_p$ is the proton mass, $\Omega_\mr{DM}$ is the DM relic density and $\Omega_b$ is the baryon density of the universe.  This relation implies a DM mass $m_{X}\simeq 5 \mbox{ GeV}$.  Though the size of this mass is phenomenologically viable, it does not directly link the DM sector to the new physics which stabilizes the weak scale.

If the $L$-violating operators which transfer the asymmetry have not decoupled as the DM becomes non-relativistic, there is a Boltzmann suppression of the DM asymmetry (see \cite{barr,harveyturner} for a more detailed discussion)
\begin{equation}
(n_X-n_{\bar{X}}) \sim (n_{\ell} - n_{\bar{\ell}})\, e^{-m_X/T_d},
\end{equation}
where $T_d$ is the temperature at which the $L$-violating operators decouple.
This implies that the DM mass can be much larger \footnote{In deriving this relation we have assumed that the universe reheated high enough for the electroweak sphalerons to be active and that they remain in equilibrium at temperatures below the electroweak phase transition.  Hence, at the sphaleron decoupling temperature we assume that the top quark and $H'$ particles do not contribute to the relevant number densities.}
\begin{equation}
m_X = \frac{45}{29}\,\frac{1}{N_X}\,\frac{f(0)}{f(m_X/T_d)}\, \frac{\Omega_\mr{DM}}{\Omega_b}\, m_p,
\label{DMmass}
\end{equation}
where $N_X$ is the number of DM families and $f(x)$ is the Boltzmann suppression factor given by
\begin{equation}
f(x)=\frac{1}{4\,\pi^2}\int_0^\infty \frac{y^2\,\mathrm{d}y}{\mathrm{cosh}^2(\frac{1}{2}\sqrt{y^2+x^2})}.
\end{equation}

The decoupling temperature, $T_d$, is naturally at the electroweak scale if the corresponding higher dimensional operators are TeV scale suppressed.  Once these $L$-violating operators decouple, the asymmetric DM density is frozen in.

Although the $L$-violating interactions have frozen out, $L$-preserving interactions are expected to remain in thermal equilibrium to lower temperatures.  This is particularly natural if the $L$-violating operators are generated by a combination of the $L$-preserving interactions and an operator which introduces a small amount of $L$-violation into the theory.   While the $L$-preserving operators may be in thermal equilibrium longer than the resulting $L$-violating interactions, they do not change the {\em relic} DM density, which will be dominantly composed of $\bar{X}$s with essentially no $X$s.

If the asymmetry in the DM persisted until today, there would be no indirect detection signal from $X-\bar{X}$ annihilation.  If, however, there is a small violation of DM number in the dark sector, as may result from a small DM Majorana mass, $X-\bar{X}$ oscillations will erase the asymmetry without reducing the relic density, giving rise to a signal for indirect detection experiments from $\bar{X}\, X \rightarrow \ell^+\, \ell^-$.   In some cases the hidden sector may be more complicated, and four lepton final states may also result, {\em e.g.} $\bar{X}\, X \rightarrow \ell^+\, \ell^-\, \ell^+\, \ell^-$.  Since this $L$-preserving interaction is expected to be stronger than the $L$-violating operator which set the asymmetry, the associated annihilation cross-section may be large enough to generate the cosmic ray positron excesses.

There are many models which exhibit the generic features described above.  The rest of the letter is devoted to an illustrative toy model which reproduces this scenario.  Consider the $L$-violating interaction (from \cite{ADM})
\begin{equation}
{\cal L}_\mr{asym} = \frac{1}{{M'}_{ij}^4} \bar{X}^2(L_i\, H)(L_j\, H) +\, \mbox{h.c.},
\label{nonSUSYLviol}
\end{equation}
where $L$ is the lepton doublet, $H$ is the SM Higgs doublet and $M'$ is a new $L$-violating mass scale.  This term mediates $\bar{X}\,\bar{X} \leftrightarrow \bar{\nu}\, \bar{\nu}$, thereby transferring the lepton asymmetry to an $X-\bar{X}$ asymmetry.  Consider in addition the $L$-preserving interaction
\begin{equation}
{\cal L}_\mr{sym} = \frac{1}{M_{ij}^2} \bar{X}\,X\, \bar{L}_i\, L_j +\, \mbox{h.c.},
\label{nonSUSYLpreserve}
\end{equation}
where $M$ is a new $L$-preserving mass scale, which mediates $\bar{X}\, X \leftrightarrow \ell^+\, \ell^-,\,\bar{\nu}\,\nu$.  A UV completion of these operators is
\begin{equation}\label{eq:UVCompletion}
{\cal L} \ni y_i\, L_i\, H'\,\bar{X} - \frac{\lambda'}{2}(H^\dagger\, H')^2+\mbox{h.c.},
\end{equation}
where $H'$ is a new Higgs doublet.  There is a $\mathbb{Z}_2$ symmetry under which $X$, $\bar{X}$ and $H'$ are charged, which is unbroken for $\langle H' \rangle = 0$.  This symmetry ensures that the lightest $\mathbb{Z}_2$ odd state, which we take to be $\bar{X}$, is stable.  Upon integrating out $H'$, the effective scale of $L$-violation (Eq.~(\ref{nonSUSYLviol})) is ${M'}_{ij}^4 = m_{H'}^4/(y_i\, y_j\, \lambda')$, and the scale of the $L$-preserving operator (Eq.~(\ref{nonSUSYLpreserve})) is ${M}_{ij}^2 = m_{H'}^2/(y_i\, y_j)$.  Also note that while the model with $N_X =1$ does not violate $L$, it does violate any two of electron number, muon number and tau number due to the first interaction in Eq.~(\ref{eq:UVCompletion}).  For weak scale parameters and assuming that $y_i = y \simeq 1$, the rate for $\mu\rightarrow\,e\,\gamma$ is $\sim 15$ orders of magnitude above the current bound.  One way to avoid this bound is to assume a hierarchy of $\mathcal{O}(10^{-8})$ between the first two generations of $y_i$ couplings.  For $N_X = 3$ the interactions are expanded to
\begin{equation}\label{eq:Lwith3XFamilies}
{\cal L } = y_{ij}\, L_i H'\, \bar{X}_j + m_X^i\, \bar{X}_i\, X_i.
\end{equation}
For a generic $y_{ij}$ matrix, the same large rates for $\mu\rightarrow e\,\gamma$ are present as describe above for $N_X=1$.  If $y_{ij} = \mr{diag}(y_1,y_2,y_3)$ in this basis (where $m_X$ is diagonal), contributions to $\mu \rightarrow e\,\gamma$ vanish.

The $\lambda'$ term is present in Eq.~(\ref{eq:UVCompletion}) to break a global $U(1)_X$, under which $X,~\bar{X}$ and $H'$ are charged so that an $X$ asymmetric operator such as Eq.~(\ref{nonSUSYLviol}) can arise.  For $M$ and $M'$ at or above the electroweak scale and $\lambda' < 1$, $(M'^2_{ij}) \gtrsim (v\,M_{ij})$, implying that the $L$-violating operators decouple first ($v\equiv \langle H \rangle$).  The annihilations through the operator in Eq.~(\ref{nonSUSYLpreserve}) (and Eq.~(\ref{Diracmass}) below) give rise to larger cross-sections than through Eq.~(\ref{nonSUSYLviol}).  The smaller cross-section from the $L$-violating operators set the DM asymmetry, and hence its relic density.

From Eq.~(\ref{DMmass}), $m_{X}/T_d \approx 5-8$ for $m_X \approx 100-1000$ GeV (note there is only logarithmic sensitivity to $m_X$).  Then using $H(T_d) = n_{\bar{X}}\,\langle \sigma_\mr{asym}\,v\rangle$ to set the $L$-violating cross-section yields $\lambda' = 2\times 10^{-4}$ for $m_X = 500$ GeV, $N_X = 1$ and $y = 1$, or equivalently $M' \simeq \ 5 \mbox{ TeV}\, (m_X/500\mbox{ GeV})^{3/8}\,N_X^{1/8}$.
For reference we include the zero temperature result for the asymmetric annihilation $\bar{X}\, \bar{X} \leftrightarrow \bar{\nu}\, \bar{\nu}$
\begin{equation}
\langle \sigma_\mr{asym}\, v \rangle = \frac{1}{16\,\pi}\,\frac{v^4\,m_X^2}{M'^8},
\label{annXsectn}
\end{equation}
which results in an $\mathcal{O}(20\,\%)$ error when calculating $M'$.

The symmetric annihilation $\bar{X}\, X \leftrightarrow \ell^+\, \ell^-,\,\bar{\nu}\,\nu$ through Eq.~(\ref{nonSUSYLpreserve}) with cross-section
\begin{equation}\label{eq:sigmaAnnSym}
\langle \sigma_\mr{sym}\, v \rangle = \frac{1}{8\,\pi}\,\frac{m_X^2}{M_{ij}^4},
\end{equation}
will typically freeze-out at a temperature lower then $T_d$.  These annihilations do not affect the relic density, which is set by the DM asymmetry.

As long as the DM density is asymmetric, there will be no indirect signals for DM in the universe today.  However, a small Majorana mass $m_M$ term,
\begin{equation}
{\cal L}_M = m_M\, \bar{X}\, \bar{X},
\end{equation}
will induce $X-\bar{X}$ oscillations which erase the DM asymmetry and give rise to $X-\bar{X}$ annihilation signals in the universe today.  For $m_X = 500$ GeV and $M = 300-600$ GeV (corresponding to $y=2-1$ and $m_{H'}= 600$ GeV), $\langle \sigma_\mr{sym} v \rangle = 10^{-23}-10^{-24} \mbox{ cm}^3/\mbox{s}$ which is the size required to generate the PAMELA and Fermi signals.

One can also generate four lepton final states in this model with only a minor modification.  For example the Dirac mass term, $m_X\, \bar{X}\, X$, could result from the vev of a new singlet scalar ($\Phi$) and the interaction
\begin{equation}
{\cal L}_X = \lambda_X\, \Phi\, \bar{X}\, X,
\label{Diracmass}
\end{equation}
where $m_X \equiv \lambda_X\,\langle \Phi\rangle$.  Assuming $\Phi$ has no direct couplings to the SM, its decays will occur exclusively to leptonic final states through a one-loop diagram.  Then the interactions in Eq.~(\ref{Diracmass}) mediate annihilations to $\bar{X}\, X \rightarrow \Phi\, \Phi\, \rightarrow \ell^+\, \ell^-\, \ell^+\, \ell^-$.  Note that we do not require kinematic restrictions to force $\Phi$ to decay to leptonic final states.

There is a cosmological restriction on the $X$ Majorana mass -- to preserve the relic density, we require that no annihilations recouple when the $X-\bar{X}$ oscillations commence.  Otherwise the relic density would be reduced to the (small) thermal value set by the symmetric processes.  Quantitatively, the symmetric ``no-recoupling'' temperature ($T_\mr{nr}$), defined by
\be
\frac{n_\mr{asym}(T_\mr{nr})}{2}\,\langle \sigma_\mr{sym}\,v \rangle = H(T_\mr{nr}),
\ee
must be greater than the temperature when oscillations begin ($T_\mr{osc}$):
\be\label{eq:norecouplingCondition}
H(T_\mr{nr}) \gtrsim H(T_\mr{osc}) \sim m_M.
\ee
For the no-recoupling relation, we have taken equal parts $\bar{X}$ and $X$ from oscillations at $T_\mr{nr}$, and $n_\mr{asym}$ is the relic DM density set by asymmetric annihilations.  Using Eq.~(\ref{DMmass}) to find $n_\mr{asym}(T_\mr{nr})$ and Eq.~(\ref{eq:sigmaAnnSym}) we find $T_\mr{nr} \simeq 0.8\,\mr{GeV}\,g_*^{-1/2}\,(10^{-23}\mbox{ cm}^3/ \mbox{s}/ \langle \sigma_a\,v \rangle)$ for $m_X = 500$ GeV.  Then Eq.~(\ref{eq:norecouplingCondition}) implies $m_M \lesssim \mathcal{O}(10^{-14}-10^{-20}\,\mr{GeV})$ for $\langle \sigma_\mr{sym}\,v\rangle \sim \mathcal{O}(10^{-26}-10^{-23}\mbox{ cm}^3/ \mbox{s})$.  This very small mass is natural since $X$ effectively carries lepton number, an unbroken global symmetry in the absence of Majorana neutrino masses.  Then the presence of Majorana neutrino masses induces an $X$ Majorana mass:
\begin{equation}\label{eq:oneLoopXMajoranaMass}
m_M \sim \frac{1}{16\, \pi^2}\,y^2\,\lambda'\, v^2\, \frac{m_{\nu}}{m_{H'^0}^2} \sim \mathcal{O}(10^{-18}\,\mathrm{GeV}),
\end{equation}
where the last relation is for the parameters described above Eq.~(\ref{annXsectn}).  This is a small enough Majorana mass that no wash out occurs for $\langle \sigma_\mr{sym} v \rangle \lesssim 10^{-24} \mbox{ cm}^3/\mbox{s}$.  Also note that since we are assuming instantaneous oscillations, even when $m_M$ is at the upper bound of the constraint implied by Eq.~(\ref{eq:norecouplingCondition}) there will only be an $\mathcal{O}(1)$ change in the DM relic density.  Thus for the symmetric annihilation cross-sections of interest here, Majorana neutrino masses are often consistent with the no-recoupling condition.  Models with mass varying neutrinos \cite{mavan} or where the neutrinos are Dirac will weaken this or eliminate this constraint.

The constraints from neutrino masses also do not apply if the $X$ Majorana mass {\em induces} Majorana neutrino masses.  If the $X$ Majorana mass results from the vev of a sub-GeV scalar field ($S$), from the interaction
\begin{equation}\label{eq:MajoranaMassFromSVev}
{\cal L}_M = \kappa_{\alpha \beta}\,S\,\bar{X}_{\alpha}\,\bar{X}_{\beta},
\end{equation}
and the scalar field only obtains a vev at $T < T_\mr{nr}$, the Majorana mass $((m_M)_{\alpha \beta} \equiv \kappa_{\alpha \beta}\,\langle S \rangle)$ can be arbitrarily large without reducing the DM number density.  In this case, the neutrino mass is generated at one-loop \cite{PerezWise}:
\begin{equation}\label{eq:neutrinoMass}
(m_\nu)_{ij} = y_{i\alpha}\, y_{j\beta} \frac{\lambda'}{16\, \pi^2}\, v^2\, \frac{(m_M)_{\alpha \beta}}{m_{H'^0}^2},
\end{equation}
where we have taken $N_X=3$.  Since one must assume that $y_{ij}$ is flavor diagonal to avoid lepton flavor violating decays, the flavor and CP violation in the neutrino sector result from the structure of the $X$ Majorana mass matrix.  The parameters $y \sim \mathcal{O}(1)$, $\lambda' \sim \mathcal{O}(10^{-4})$ and $m_{H'^0} \sim \mathcal{O}(600\,\mr{GeV})$ require $m_M \sim \mathcal{O}(10^{-5}\,\mr{GeV})$ to achieve $m_{\nu}\sim \mathcal{O}(10^{-2}\,\mr{eV})$.  The off-diagonal entries in $m_M$ lead to $\mu \rightarrow e\,\gamma$ but for these parameters the constraint is satisfied.

One might worry that the interaction in Eq.~(\ref{eq:MajoranaMassFromSVev}) could wash out the $X$ asymmetry through, {\em e.g.}, $\bar{X}\,\bar{X} \leftrightarrow S\,S$ processes.  The $X$ asymmetry is safe from wash out provided this process decouples above $T_d$, which happens for small $U(1)_X$ violation, $\kappa \lesssim \mathcal{O}(10^{-3})$.  The phase transition to the vacuum with a non-zero vev for $S$ obtains if either the temperature drops below the critical temperature associated with the $S$ potential or the $S$ particles decay.  $S$ decays to two neutrinos via a one-loop diagram with rate $\Gamma_{S-\mr{decay}} \sim \mathcal{O}(10^{-22}\,\mr{GeV})$ for the parameters discussed above and $m_S \simeq 10 \mbox{ MeV}$.  The decay happens just after $S$ becomes non-relativistic but before big bang nucleosynthesis, avoiding any cosmological problems. 

This model does not possess any DM-nucleon couplings at tree-level.  However, the operator in Eq.~(\ref{nonSUSYLpreserve}) induces an effective magnetic dipole moment for the DM when coupling a photon to the lepton loop.  This leads to a direct detection cross-section for $X$ scattering off of a nucleon (see \cite{ADM} and the references therein for details)
\begin{eqnarray}
\sigma_{dd} \simeq 2 \times 10^{-46}\,\mathrm{cm}^2\, \left(\frac{Z/A}{0.4}\right)^2\left(\frac{600\,\mathrm{GeV}}{m_{H'^{\pm}}/y}\right)^4. 
\end{eqnarray}
This will be a signal for the next generation of experiments.

To conclude, relating the lepton asymmetry to the DM density implies a novel mechanism for obtaining both leptophilic DM and a separation between the freeze-out and present day annihilation cross-sections.  In these models, $L$-violating operators which transfer the lepton asymmetry set the DM density, while related $L$-preserving operators set the rates for annihilation in indirect detection experiments (such as PAMELA and Fermi).  The smaller $L$-violating cross-sections set the relic density, while allowing for large cross-sections for indirect detection experiments through the $L$-preserving operators.  If DM of this type is responsible for the cosmic ray anomalies, then it will be observed in the next generation of direct detection experiments.  Non-minimal versions of the model can generate the SM neutrino masses and mixings at one-loop.  Such classes of Asymmetric Dark Matter will continue to be important for both model building and experimental searches for DM in the galaxy today.

\bigskip

This work has been supported by the US Department of Energy, including the grant DE-FG02-95ER40896 (KMZ) and the National Science Foundation, including the NSF CAREER Grant NSF-PHY-0743315 (TC).  We thank Roni Harnik, Markus Luty, Ann Nelson, Frank Petriello, and Aaron Pierce for helpful discussions, and Graham Kribs for hosting the Unusual Dark Matter Workshop at the University of Oregon (under DOE contract DE-FG02-96ER40969) where some of this work was developed.

\end{document}